\documentclass[prd,nofootinbib,12pt]{revtex4}

\usepackage{lineno,hyperref}
\usepackage{amsfonts,amssymb,amsmath}
\usepackage{epsfig}
\usepackage{mathrsfs}
\usepackage{amssymb}
\usepackage{graphicx}
\usepackage{amsmath}
\usepackage{graphicx}
\usepackage{amsmath}
\usepackage{xcolor}
\usepackage{color}
\usepackage{appendix}

\begin{document}

\title{Inflationary Krylov complexity}

\author{Tao Li$^{1}$}
\author{Lei-Hua Liu$^{1}$}
\email{liuleihua8899@hotmail.com}
\affiliation{$^1$Department of Physics, College of Physics, Mechanical and Electrical Engineering, Jishou University, Jishou 416000, China}

\begin{abstract}

In this work, we have systematically investigated the Krylov complexity of curvature perturbation for
the modified dispersion relation in inflation, using the algorithm in closed system and open system. Since many quantum gravitational frameworks could lead to this kind of modified dispersion relation, our analysis could be applied to the string cosmology, loop gravity, $\it e.t.c$. Following the Lanczos algorithm, we find the very early universe is an infinite, many-body, and maximal chaotic system. Our numerics shows that the Lanczos coefficient and Lyapunov index of the standard dispersion relation are mainly determined by the scale factor. As for the modified case, it is nearly determined by the momentum. In a method of the closed system, we discover that the Krylov complexity will show irregular oscillation before the horizon exits. The modified case will present faster growth after the horizon exists. Since the whole universe is an open system, the approach of an open system is more realistic and reliable. Then, we construct the exact wave function which is very robust only requiring the Lanczos coefficient proportional to $n$ (main quantum number). Based on it, we find the Krylov complexity and Krylov entropy could nicely recover in the case of a closed system under the weak dissipative approximation, in which our analysis shows that the evolution of Krylov complexity will not be the same with the original situation. We also find the inflationary period is a strong dissipative system. Meanwhile, our numerics clearly shows the Krylov complexity will grow during the whole inflationary period. But for the small scales, there will be a peak after the horizon exits. Our analysis reveals that the dramatic change in background (inflation) will significantly impact the evolution of Krylov complexity. Since the curvature perturbation will transit from the quantum level to the classical level. We could expect that the decoherence will highly impact the Krylov complexity during inflation.

\end{abstract}

\maketitle

%%%%%%%%%%%%%%%%%%%%%%%%%%%%%%%%%%%%%%%%%%%%%%%%%%%%%
%%%%%%%%%%%%%%%%  I N T R O D U C T I O N  %%%%%%%%%%%%%%%%%%%%%%%%%
%%%%%%%%%%%%%%%%%%%%%%%%%%%%%%%%%%%%%%%%%%%%%%%%%%%%%

\section{introduction}
\label{introduction}
In light of the holographic principle, spacetime emerges from the quantum entanglement \cite{VanRaamsdonk:2010pw}. A more striking concept was proposed which claimed that Einstein-Rosen bridge (ER)= Einstein–Prodolsky–Rosen pair (EPR) \cite{Maldacena:2013xja}. Thereafter, Ref. \cite{Stanford:2014jda} discovered that the boundary conformal field theory (CFT) reaches the thermal equilibrium within a short temporal interval, however, the evolution of the corresponding wormhole will take much longer time compared with CFT. To solve this issue, the so-called complexity was proposed \cite{Hartman:2013qma,Liu:2013iza}. In the holographic framework, quantum entanglement produces the spacetime, and the evolution of spacetime is responsed by the complexity. 

The holographic principle needs the conformal symmetry for the quantum field theory. Meanwhile, with the development of complexity in high-energy physics, it is natural to investigate the complexity in quantum systems. Ref. \cite{Aaronson:2016vto} defined complexity as the minimal operations for a certain task, namely also dubbed as the computational complexity. Afterward, there were two main methods for calculating the complexity: (a) The geometric method of Nielsen  {\it et al} \cite{Nielsen2006AGA,Nielsen2006QuantumCA,Dowling2008TheGO}. (b) The method of "Fubini-Study" distance on the related quantum states \cite{Chapman:2017rqy}. To be more precise, these two main methods, especially the Nielson's geometric method, are aimed to investigate the so-called circuit complexity, what they investigate is the minimal number of quantum operations from a reference state to a target state. The third method was called the Krylov complexity, Ref. \cite{Parker:2018yvk} introduced the Krylov complexity as the measurement of operator growth with respect to Krylov basis, in which they proposed a universal operator growth in the chaotic system. Compared with the previous two methods, the definition of Krylov complexity is free of ambiguities since its definition is unique. For the other two methods, the complexity depends on the choice of quantum gates (the geometric method) or the metric (the method of "Fubini-Study" distance).  Once the quantum state is defined, one can construct the Krylov space in light of the Lanczos algorithm \cite{viswanath1994recursio}, and one can calculate the Krylov complexity, Lanczos coefficient, and its corresponding entropy. Ref. \cite{Aguilar-Gutierrez:2023nyk} has shown that  one cannot make the Krylov and Nielsen notions of complexity compatible with each other, whereas under certain assumptions, one can allow for Krylov complexity and "Fuibini-Study" length to be proportional to each other \cite{Caputa:2021sib}.

Ref. \cite{Muck:2022xfc} systematically developed the method for calculating the Krylov complexity in various orthogonal polynomials dubbed as the inner product. The Krylov complexity can be applied for the SKY model \cite{Parker:2018yvk,Rabinovici:2020ryf,Jian:2020qpp,He:2022ryk}. One can also define a thermalized Krylov complexity \cite{Kar:2021nbm}. Even the chaos can be understood as the delocalization in Krylov space \cite{Dymarsky:2019elm}. By contrast, Ref. \cite{Rabinovici:2021qqt} proposed the Krylov complexity is suppressed by the finite size which is in the integrable models. For further tests and numerical calculations, it can be implemented into the generalized coherent states \cite{Caputa:2021sib,Patramanis:2021lkx},  the Ising and 
Heisenberg models \cite{Cao:2020zls,Trigueros:2021rwj,Heveling:2022hth}, CFT \cite{Dymarsky:2021bjq,Caputa:2021ori},  topological phases of matter \cite{Caputa:2022eye}, 
integrable models with saddle-point dominated scrambling \cite{Bhattacharjee:2022vlt}, cosmological complexity \cite{Adhikari:2022oxr} $\it e.t.c$. The very recent developments can be found in Refs. \cite{Erdmenger:2023wjg,Patramanis:2023cwz,Fan:2023ohh,Hashimoto:2023swv,Vasli:2023syq,Domingo:2023kjr,Gill:2023umm,Bhattacharjee:2023uwx,Adhikari:2022whf}. Especially, Ref. \cite{Camargo:2023eev}  involves examining the Krylov operator complexity within the simple quantum billiard model, which can be considered as the extension to finite temperature of \cite{Hashimoto:2023swv}. Ref. \cite{Huh:2023jxt} also has verified all the universal features of spread complexity within saddle dominated scrambling ("integrable" systems). 

One of the main motivations is to explore the Krylov complexity in the very early universe since the energy of the universe is not conserved due to the symmetry broken along the temporal part. It is naturally considering our universe as an open system. There are some attempts for using the complexity in cosmology, see Refs \cite{Adhikari:2021ked,Choudhury:2020hil}. 
To be more precise, one could consider that the universe will exchange information with the environment (outer of the universe). Following Refs. \cite{Bhattacharya:2023zqt,Bhattacharjee:2022lzy}, we will write the Hamilton of single field inflationary model with modified dispersion relation into one part that governs the closed and unitary system, another part will response the information exchanging with the environment. This kind of treatment can be considered as the Markovian environment in terms of Lindbladian evolution \cite{Lindblad:1975ef,Gorini:1975nb}. It should be noticed that the whole Hamilton is non-Hermitian under this Lindbladian evolution. From Ref. \cite{Socorro:2022aoz}, it indicates that the gravity system (k-essence) is non-Hermitian. Therefore, it is quite natural to use the approach of open system.

As for the application of complexity in cosmology, we already investigated the complexity in the non-trivial sound speed model and k-essence \cite{Liu:2021nzx,Li:2021kfq}. The evolution of complexity is similar during the inflationary period and after inflation. However, Ref. \cite{Li:2023ekd} discovered that the modified dispersion relation will lead to the various patterns of the complexity evolution in inflation and the late period. The modified dispersion in inflation was proposed by Ref. \cite{Cai:2009hc}, various frameworks of quantum gravity will lead to this kind of modified dispersion relation \cite{Armendariz-Picon:2003jjq,Armendariz-Picon:2006vgx,Magueijo:2008sx,Martin:2000xs,Arkani-Hamed:2003pdi,Bojowald:2006zb,Jacobson:2000gw,Cai:2007gs}, meanwhile it has many phenomenological implications \cite{Cai:2009in,Li:2009rt,Cai:2009zp,Cai:2012yf,Cai:2018tuh,Zheng:2017qfs,Chen:2017dhi,Bianco:2016yib,Pan:2015tza}, such as string cosmology, DBI inflation, loop gravitational cosmology $\it e.t.c$. 
Thus, our investigation is valid for these kinds of quantum gravitational frameworks. 

We organize the paper as follows. In Sec. \ref{Krylov complexity}, we will give a brief introduction to Krylov complexity and the Lanczos algorithm, which is the method for dealing with closed system. In Sec. \ref{modified dispersion relation and inflation}, some fundamental knowledge about the modified dispersion relation and inflation will be given to readers. Sec. \ref{krylov complexity of modified dispersion relation} will give the calculation of Krylov complexity, its corresponding Krylov entropy and discuss some properties in inflation. In Sec. \ref{krylov in open}, we will generalize the Lanczos algorithm into the open system and discuss the general properties of Krylov complexity in a weak dissipative system. Then, we will construct an exact wave function for calculating the Krylov complexity and Krylov entropy. Finally, the conclusion and outlook will be given in Sec. \ref{conclusion}.

%%%%%%%%%%%%%%%%%%%%%%%%%%%%%%%%%%%%%%%%%%%%%%%%%%%%%%%%
%%%%%%%%%%%%%%%%%%%    T H E     M O D E L   %%%%%%%%%%%%%%%%%%%%%%%%%%
%%%%%%%%%%%%%%%%%%%%%%%%%%%%%%%%%%%%%%%%%%%%%%%%%%%%%%%%

\section{Krylov complexity and Lanczos algorithm}
\label{Krylov complexity}

In this section, we will briefly introduce the concept of Krylov complexity and its corresponding Lanczos algorithm. It is convenient for working in the Heisenberg picture, first, we could have 
\begin{equation}
\begin{split}
\partial _{t}\mathcal{O} (t)=i[H,\mathcal{O} (t)]
\end{split}   
\label{operator o}
\end{equation}
where $\mathcal{O}(t)$ is the operator depending on time and $H$ is the Hamilton. Its corresponding solution could be written by
\begin{equation}
\begin{split}
\mathcal{O} (t)=e^{iHt}\mathcal{O} e^{-iHt}
\end{split}
\end{equation}
Defining Liouvillian super-operator $\mathcal{L}_{X}$ as $\mathcal{L}_{X}Y=[X,Y]$ where $[,]$ is the commutator as shown in Eq. \eqref{operator o}. Under this operator, the general operator can be written by 
\begin{equation}
\begin{split}
\mathcal{O} (t)=e^{i\mathcal{L}t}\mathcal{O} =\sum_{n=0}^{\infty } \frac{(it)^{n}}{n!} \mathcal{L}^{n}\mathcal{O} (0)=\sum_{n=0}^{\infty } \frac{(it)^{n}}{n!} \tilde{\mathcal{O} }_{n}  
\end{split}
\label{wave function of O}
\end{equation}
where $\mathcal{O}(t)$ can be regarded as the wave function, $\mathcal{L}$ is the Liouvillian super-operator and $ \mathcal{L}^{n}\mathcal{O} =\tilde{\mathcal{O} }_{n} =[H,\tilde{\mathcal{O}}_{n-1}]$ which can form the Hilbert space and could expand as follows, 
\begin{equation}
\begin{split}
\mathcal{O}\equiv |{\tilde{\mathcal{O}} })  ,\mathcal{L}^{1}\mathcal{O}\equiv |{\tilde{\mathcal{O}} }_{1})  ,\mathcal{L}^{2}\mathcal{O}\equiv |{\tilde{\mathcal{O}} }_{2})  ,\mathcal{L}^{3}\mathcal{O}\equiv |{\tilde{\mathcal{O} } }_{3}) ...      
\end{split}
\label{basis of O}
\end{equation}
however, these bases are not orthogonal to each other. Then, we could implement the Lanczos algorithm to construct an orthogonal Krylov basis, in which the first two operators are 
\begin{equation}
\begin{split}
\mathcal{O}_{0}= |{\tilde{\mathcal{O}} }_{0}) =\mathcal{O},~~~ |\mathcal{O}_{1})=b_{1}^{-1}\mathcal{L}|\mathcal{O}_{0}) 
\end{split}
\label{krylov basis}
\end{equation}
where $b_{1}=\sqrt{(\tilde{\mathcal{O}}_{0}\mathcal{L}|\mathcal{L}\tilde{\mathcal{O}}_{0})  }$ desribing the normalized vector. The other orthogonal basis can be formed by the following iterative relation, 
\begin{equation}
\begin{split}
|A_{n})=\mathcal{L}|\mathcal{O}_{n-1})-b_{n-1} |\mathcal{O}_{n-2})
\end{split}
\label{An}
\end{equation}
with
\begin{equation}
\begin{split}
|\mathcal{O}_{n})=b_{n}^{-1} |A_{n}),~~b_{n}=\sqrt{(A_{n}|A_{n})}
\end{split}
\label{On}
\end{equation}
If $b_n=0$, this iterative relation will stop and that will produce the finite orthogonal Krylov basis, and a set of ${b_n}$ is called the Lanczos coefficients. Thereafter, we could expand the \eqref{wave function of O} as 
\begin{equation}
\begin{split}
\mathcal{O}(t)=e^{i\mathcal{L}t}\mathcal{O}=\sum_{n=0}^{\infty } (i)^{n} \phi_{n}(t)|\mathcal{O}_{n})  
\end{split}
\label{Ot}
\end{equation}
where $\phi_n$ is the wave function and it satisfies with the normalized condition $\sum_{n}\left | \phi_{n} \right |^{2}=1$. Substitute Eq. \eqref{Ot} into the Schr$\ddot{o}$dinger equation, one could obtain 
\begin{equation}
\begin{split}
\partial _{t}\phi_{n}(t)=b_{n}\phi_{n-1}-b_{n-1}\phi_{n+1}. 
\end{split}
\label{eom of phi}
\end{equation}
Afterward, one can define the Krylov complexity in terms of $\phi_n$ as follows 
 \begin{equation}
 \begin{split}
 K=\sum_{n=1}^{k}n\left | \phi_{n} \right | ^{2},
 \label{krylov complexity}
 \end{split}
 \end{equation}
where $k$ is the dimension of Krylov basis $\mathcal{O}_n$. In addtion, Ref. \cite{Parker:2018yvk}
has proposed that the Krylov complexity of the many-body system in the thermodynamic limit, the Lanczos coefficient $b_n$ will be conjectured as follows, 
\begin{equation}
b_{n}\le \alpha n+\eta ,
\label{approximated growth}
\end{equation}
where $\alpha$ and $\eta$ are the information about the chaotic characteristics of the many-body system. When Eq. \eqref{approximated growth} becomes $b_n=\alpha n+\eta$, it arrives at the maximal growth which presents an ideal chaotic system confirmed by \cite{Bhattacharya:2022gbz}. And the value of $n$ becomes large, $b_n$ will also be proportional to $n$. In the latter investigation, we will show that $b_n$ of our case is perfectly proportional to $n$, thus the Lyapunov index $\lambda$ has the relation denoting the exponential growth of the Krylov operator complexity as follows,
\begin{equation}
\lambda=2\alpha. 
\label{lyapunov index}
\end{equation}
In the latter investigations, the key point is that we will calculate Krylov complexity, Lanczos coefficient, and Lyapunov index for the modified dispersion relation in inflation \cite{Cai:2009hc}. 

\section{Revisit the modified dispersion relation and inflation}
\label{modified dispersion relation and inflation}
In this section, we will give brief information on modified dispersion relation in inflation and some basic knowledge of inflation.

\subsection{Modified dispersion relation}
\label{modified dispersion relation}

Under the Friedman–Lemaitre––Robertson Walker background metric,
\begin{equation}
ds^{2} =a(\eta ) ^{2}(-d\eta ^{2}+d\vec{x} ^{2} ),
\label{FRW metric}
\end{equation}
where $\Vec{x}=(x,y,z)$ is the spatial vector and $a(\eta)$ is the scale factor in conformal time. The curvature perturbation is inside the Hubble radius. As for the expansion of the universe, the wavelength is also becoming larger, therefore this kind of curvature perturbation will re-enter the Hubble radius. This physical process can be characterized by which the wavelength $k/a$ is longer than $1/H$. Thus, the conformal time is more convenient. Under metric \eqref{FRW metric}, one can define the curvature of metric as follows, 
\begin{equation}
ds^{2} =a(\eta ) ^{2}\left ( -(1+\psi(\eta,x)  d\eta ^{2}+(1-\psi(\eta,x)d\vec{x}^{2}  \right ),
\label{perturbation of metric}
\end{equation}
where $\psi(\eta,x)$ is the perturbation of metric and meanwhile we can define the perturbation of scalar field as $\phi (x_\mu )=\phi_0(\eta )+\delta \phi (x_\mu )$. Being armed with these two perturbations, the perturbated action can be derived as follows,
\begin{equation}
S=\frac{1}{2} \int dtd^3xa^3\frac{\dot{\phi } }{H^2}\left [ \mathcal{\dot{R}}- \frac{1}{a} (\partial _i\mathcal{R} )^2 \right ],
\label{perturbated action}
\end{equation}
where \(H=\frac{\dot{a}}{a}\), \(\mathcal{R}=\psi+\frac{{H} }{\phi _0}\delta\phi\) , $z=\sqrt{2\epsilon}a$, $\epsilon=-\frac{\dot{H}}{H^2}=1-\frac{{\mathcal{H}}'}{\mathcal{H}^2}$. Action \eqref{perturbated action} can be written in terms of the Mukhanov variable \(v =z\mathcal{R}\),
\begin{equation}
S=\frac{1}{2}\int d\eta d^3x\left [ {v}'^2-(\partial _i v )^2+\big(\frac{z'}{z}\big)^2v^2-2\frac{ {z}'}{z}{v}'v  \right ].
\label{standard action}
\end{equation}
This action is of the standard dispersion relation. As for the modified dispersion relation, it is introduced by \cite{Cai:2009hc} whose corresponding action reads as 
\begin{equation}
\begin{split}
S=\int d^3x d\eta L= \frac{1}{2}\int d\eta d^3x\left [ {v}'^2-f_{ph}^2(\partial _i v)^2+\big(\frac{z'}{z}\big)^2v^2-2\frac{ {z}'}{z}{v}'v  \right ],
\label{action of vi}
\end{split}
\end{equation}
where 
\begin{equation}
\begin{cases}
& \text{ if }(k_{ph})>M; f_{ph}=(\frac{k_{ph}}{M})^{\alpha }  \\
& \text{ if }(k_{ph})\le M; f=1
\end{cases}
\label{modified dispersion relationn}
\end{equation}
where $k_{ph}=\frac{k}{a}$ denotes the physical wave number. 
The value of $\alpha$ could determine various kinds of quantum gravitational framework and $M$ is some certain energy scale that is different in various quantum gravitational models. Then, according to the values of $\alpha$, it can be classified into four cases:

$(a)$. The standard dispersion relation corresponding to $\alpha=0$ that is also the standard inflationary model \cite{Guth:1980zm}, which belongs to the IR region.

$(b)$. As $0<\alpha<2$, it also belongs to the UV region.

$(c)$. As $\alpha=2$, it is a special kind of Horv$\check{a}$-Lifshitz cosmology \cite{Kiritsis:2009sh,Calcagni:2009ar}.

$(d)$. As $\alpha>2$, it also belongs to the UV region. In order to construct the Hamiltonian operator, therefore one could define the canonical momentum,
\begin{equation}
\pi(\eta,\vec{x})=\frac{\delta L}{\delta{v}'(\eta,\vec{x})} ={v}'-\frac{{z}'}{z}v.
\end{equation}
It will lead to the Hamilton by means of $ H=\int d^3xd\eta(\pi v'-\mathcal{L} )$, 
\begin{equation}   
H= \frac{1}{2}\int  d^3x d\eta \left [ {\pi}^2+f^2(\partial _i v)^2+\frac{ {z}'}{z}(\pi v+v\pi) \right ].
\label{hamilton}
\end{equation}
Using the Fourier decomposition as follows, 
\begin{equation}
\hat{v} (\eta,\vec{x})=\int \frac{d^3k}{(2\pi)^{3/2}} \sqrt{\frac{1}{2k}}(\hat{c }_{-\vec{k} }^{\dagger}v_{\vec {k}}^{\ast }(\eta )e^{-i\vec{k\cdot }\vec{x}}+ {c_{\vec k}}v_{\vec k}e^{i\vec{k\cdot }\vec{x}}),
\end{equation}
\begin{equation}
\hat{\pi} (\eta,\vec{x})=i\int \frac{d^3k}{(2\pi)^{3/2}}\sqrt{\frac{k}{2}}(\hat{c }_{-\vec{k}}^{\dagger}u_{\vec k}^{\ast }(\eta )e^{-i\vec{k\cdot }\vec{x}}-\hat{c}_{\vec k}u_{\vec k}e^{i\vec{k\cdot }\vec{x}}).
\end{equation}
 where $\hat{c }_{-\vec{k} }^{\dagger}$ and $\hat{c}_{\vec k}$ represent the creation and annihilation operators, respectively. Being armed with these two Fourier modes, the Hamilton \eqref{hamilton} will become as follows 
\begin{equation}
\begin{split}
\hat{H}=\int{d^{3}k}\hat{H}_{k}=&\int{d^{3}k}\bigg[\frac{k}{2}(f^{2}+1)\hat{c }_{-\vec{k} }^{\dagger}\hat{c}_{-\vec{k}}+\frac{k}{2}(f^{2}+1)\hat{c}_{\vec k}\hat{c}_{\vec k}^{\dagger }\\ &+(\frac{k}{2}(f^{2}-1)+i\frac{z{}'}{z})\hat{c}_{\vec k}^{\dagger }\hat{c}_{-\vec{k} }^{\dagger }\\ &+(\frac{k}{2}(f^{2}-1)-i\frac{z{}'}{z})\hat{c}_{\vec k}\hat{c}_{-\vec{k} }\bigg].
\end{split}
\label{standard hamilton}
\end{equation}
 For these two modes, we have $[\hat c_k,\hat c_{-k}]=[\hat c_k,\hat c^\dagger_{-k}]=[\hat c^\dagger_{k},\hat c_{-k}]=[\hat c^\dagger_{k},\hat c^\dagger_{-k}]=0$, $[\hat c_k,\hat{c}_k^\dagger]=[\hat c_{-k},\hat{c}_{-k}^\dagger]=1$ ($[a,b]=ab-ba$) and making use of $(AB)^\dagger=B^\dagger A^\dagger$ ($A$ and $B$ are operators), it is straightforwardly proving that $H^\dagger=H$. 

\subsection{Basics of inflation}
\label{basics of inflation}

Before the analysis of Krylov complexity for the modified dispersion relation, we need to give some basic knowledge of inflation for the later numerics and investigations. As we discussed, the action \eqref{standard action} describes the single field inflation. Before the big bang, there is an exponential expansion from $10^{-36}$ sec to $10^{-32}$ sec, where this expansion behavior is represented by scale factor $a(\eta)$ in metric \eqref{FRW metric}. During inflation, the scale factor is $a=\exp[Ht]$ where $H=\frac{\dot{a}}{a}$ ($\dot{a}=\frac{da}{dt}$) is nearly a constant. By using $dt=ad\eta$, we will have $\mathcal{H}=a'/a=aH$ and $\eta=-1/(aH)$ which we could assume that H is constant for simplicity. During the inflation, the scale of the universe is very small, and the wavelength is short (very high energy) but it is still longer compared with the scale of the universe. When the universe experiences exponential expansion, the scale of the universe will be larger than the wavelength which imprints the observational effects on the event horizon (the boundary of the universe), which is the horizon exits. There is an essential result that will happen due to this horizon exits, where the curvature perturbation will transit from the quantum level to the classical level caused by the quantum decoherence. During this process, the energy will decrease dramatically representing by $k$.

The previous method for calculating the Krylov complexity is based on the static and flat spacetime, such as the method we will implement via Ref. \cite{Parker:2018yvk}. However, the inflation will lead to the exponential expansion of the background (the universe itself), meanwhile, the decoherence will also occur during the inflationary process, such as the minimal decoherence \cite{Burgess:2022nwu}, which will lose the information. Due to the extremal condition of inflation, we could expect the inflation and decoherence will give a new perspective on the Krylov complexity.

\section{Krylov complexity of modified dispersion relation in closed system's approach}
\label{krylov complexity of modified dispersion relation}
In this section, we will compute the Krylov complexity in a method of closed system followed by \cite{Adhikari:2022oxr}. 
The quantum state is a two-mode squeezed state as we previously discussed in \cite{Liu:2021nzx,Li:2021kfq}.

\subsection{The evolution of $r_k$}

For obtaining its wave function in Fock space, we need the following unitary operator 
\begin{equation}
\mathcal{U}_{k}=  \hat{\mathcal{S}}_{k}(r_{k},\phi_{k})\hat{\mathcal{R}}_{k}(\theta_{k}),
\label{unitary operator}
\end{equation}
with 
\begin{equation}
\hat{\mathcal{R} }_{\vec{k}}(\theta _{k})=\exp[-i\theta _{k(\eta)}(\hat{c}_{\vec {k}}\hat{c}_{\vec{k}}^{\dagger }+\hat{c}_{-\vec{ k}}^{\dagger }\hat{c}_{-\vec{ k}})]
\label{rotation operator}
\end{equation}
\begin{equation}
\hat{\mathcal{S}}_{\vec{k}}(r_{k},\phi_{k})=\exp[r_{k}(\eta)(e^{-2i\phi_{k}(\eta)}\hat{c}_{\vec{k}}\hat{c}_{-\vec{k}}-e^{2i\phi_{k}(\eta)}\hat{c}^{\dagger}_{-\vec{k}}\hat{c}^{\dagger}_{\vec{k}})]
\label{squeezed operator}
\end{equation}
where $\hat{\mathcal{R} }_{\vec{k}}(\theta _{k})$ is the rotation operator and $\hat{\mathcal{S}}_{\vec{k}}(r_{k},\phi_{k})$ is the squeezing operator. The rotation operator only contributes to the phase of the wave function which can be neglected. Using $\hat{\mathcal{S}}_{\vec{k}}(r_{k},\phi_{k})$ to act on the vacuum state, 
\begin{equation}
\left | \psi  \right \rangle_{sq} =\frac{1}{\cosh(r_k)}\sum_{n=0}^{\infty }(-1)^{n}e^{2in\phi_{k}}\tanh^{n}r_{k} \left | n; n \right \rangle _{\vec{k},-\vec{k}}. 
\label{wave function of squeezed state}
\end{equation}
Combine with Hamiltonian operator and Schr$\ddot{o}$dinger equation $i\frac{d}{d\eta} \left | \psi  \right \rangle _{sq}=\hat{H}_{k}\left | \psi  \right \rangle_{sq} $, one can obtain the evolution equation of $r_k$ and $\phi_k$, respectively,
\begin{equation}
\begin{split}
& -\frac{dr_{k}}{d\eta}=\frac{k}{2}(f^{2}-1)\sin(2\phi_{k})+\frac{{z}' }{z} \cos(2\phi_{k})\\&\frac{d\phi_{k}}{d\eta}=\frac{k}{2}(f^{2}+1)-\frac{k}{2}(f^{2}-1)\cos(2\phi_{k})\coth(2r_{k})+\frac{{z}' }{z}\sin(2\phi_{k})\coth(2r_{k})
\end{split}
\label{rk and phik}
\end{equation}
 If we approximate $\epsilon$ to be a constant, then we can get $\frac{{z}' }{z} =\frac{{a}' }{a}$. We also define $y=\log_{10}{a} $ and $ \eta=-\frac{1}{aH}$. Eq. \eqref{rk and phik} will become as
 follows,
 \begin{equation}
 \begin{split}
 -\frac{10^yH_0}{\ln10}\frac{dr_k}{dy}=\frac{k}{2}\bigg[(\frac{k_{ph}}{M})^{2\alpha}-1\bigg]\sin(2\phi _{k})+aH_0\cos(2\phi _{k})
 \end{split}
 \end{equation}
 \begin{equation}
 \begin{split}
 \frac{10^yH_0}{\ln10}\frac{d\phi_{k}}{dy} =\frac{k}{2}\bigg[(\frac{k_{ph}}{M})^{2\alpha}+1\bigg]+aH_0\sin(2\phi _{k})\coth(2r_{k})-\frac{k}{2}\bigg[(\frac{k_{ph}}{M})^{2\alpha}-1\bigg]\cos(2\phi _{k})\coth(2r_{k})
 \end{split}
 \end{equation}
 Until present, all of the calculations is the same with Ref. \cite{Li:2023ekd}. In the latter analysis, we will show that the Krylov complexity only depends on $r_k$. Thus, we only list the numerics of $r_k$. 
 \begin{figure}
 	\centering
 	\includegraphics[width=0.8\linewidth]{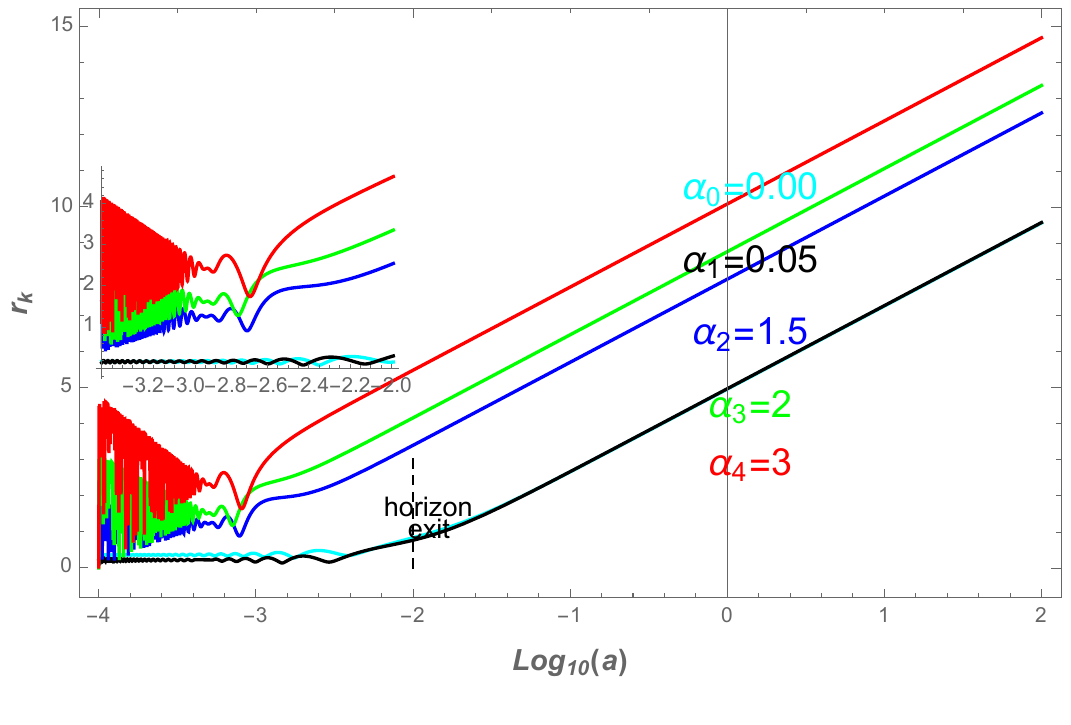}
 	\caption{The numerical solutions of $r_k(\eta)$ in terms of \(\log_{10}{a}\) with $ \alpha = 0$,  $ \alpha = 0.05$, $ \alpha = 1.5$, $ \alpha = 2$, and $ \alpha = 3$. Our plots adopt $ H_{0} = 1$ and $k=0.01$. }
 	\label{fig:rk}
 \end{figure}
Fig. \ref{fig:rk} clearly shows that there is damping behavior as $\alpha>1$ which is different from the standard dispersion relation. The more details can be found in Ref. \cite{Li:2023ekd}.

\subsection{Krylov complexity,  Lanczos coefficient and Lyapunov index}
Being armed with a numerics of $r_k$, one can investigate the complexity Lanczos coefficient and the Lyapunov index. The two-mode squeezed state wave function can be written in terms of Krylov basis as follows, 
\begin{equation}
\begin{split}
\mathcal{O}(\eta)=\sum_{n} i^{n}\phi_{n}(\eta)|{\mathcal{O}}_{n})=\frac{1}{\cosh r_{k}}\sum_{n=0}^{\infty }(-1)^{n}e^{2in\phi_{k}}\tanh^{n}r_{k} \left | n; n \right \rangle _{\vec{k},-\vec{k}}.
\end{split}
\label{wave function in Krylov basis}
\end{equation}
In our case, $|{\mathcal{O}}_{n})=\left| n; n \right \rangle_{\vec{k},-\vec{k}}$ and the operator wave function can be written by 
\begin{equation}
\phi_{n}=\frac{e^{2in\phi_{k}}\tanh^{n}r_{k}}{\cosh r_{k}}.
\label{operator wave function}
\end{equation}
Lanczos coefficient can be obtained by following the recursion relation using the tridiagonal form of the Liouvillian superoperator, 
\begin{equation}
\mathcal{L}|\mathcal{O}_{n})=b_{n+1}|\mathcal{O}_{n+1})+b_{n}|\mathcal{O}_{n-1}),
\label{recursion relation}
\end{equation}
where the Liouvillian superoperator is usually constructed by the creation and annihilation operators. In our case, it is corresponding to the Hamilton \eqref{standard hamilton}. Through a simple analysis, one can find only the following part of Hamilton will contribute to the recursion relation \eqref{recursion relation}, 
\begin{equation}
\mathcal{H}_{closed}= \bigg[\frac{k}{2}(f^{2}-1)+i\frac{a{}'}{a}\bigg]\hat{c}_{k}^{\dagger }\hat{c}_{-k}^{\dagger }+\bigg[\frac{k}{2}(f^{2}-1)-i\frac{a{}'}{a}\bigg]\hat{c_{k}}\hat{c}_{-k},
\label{closed part of hamilton}
\end{equation}
where $\mathcal{H}$ denotes the part of closed system for Hamilton. 
Acting this Hamilton on the recursion relation, one can accordingly obtain that 
\begin{equation}
\begin{split}
\mathcal{H}_{closed}|\mathcal{O}_{n})=(n+1)\bigg[\frac{k}{2}(f^{2}-1)+i\frac{a{}'}{a}\bigg]|\mathcal{O}_{n+1})+n\bigg[\frac{k}{2}(f^{2}-1)-i\frac{a{}'}{a}\bigg]|\mathcal{O}_{n-1})  
\end{split}
\label{closed hamilton}
\end{equation}
According to the definition of $b_n=\sqrt{<A|A>}$ via Eq. \eqref{An} and its detailed definition can be found in \cite{Muck:2022xfc}, it can be explicitly written as 
\begin{equation}
	b_n=n\sqrt{\bigg(\frac{k}{2}(f^{2}-1)\bigg)^{2}+\bigg(\frac{{a}'}{a}\bigg)^{2}},
	\label{Lancozs coefficient}
\end{equation}
where it clearly indicates that the inflationary universe is an ideal chaotic system. It should be noted that the the Lanczos coefficient can also be determined by the interaction parameter \cite{Bhattacharyya:2023dhp}. One can derive the corresponding complexity based on \eqref{krylov complexity},
\begin{equation}
\begin{split}
K=\sum_{n}n\left | \phi_{n} \right | ^{2}=\sum_{n=0}^{\infty} n\frac{\tanh^{2n}r_{k}}{\cosh^{2}r_{k}} =\sinh^{2}r_{k} 
\end{split}
\label{Krylov complexity1}
\end{equation}
where we have used $\sum_{m=0}^{\infty} mz^{m} =\frac{z}{(1-z)^{2}}$. It should be pointed out that the two-mode squeezed state has the maximum complexity growth bound with the $SL (2, R)$ symmetric group structure \cite{Hornedal:2022pkc}. In light of \eqref{lyapunov index} and $\alpha=\frac{b_n}{n}$, it straightforwardly has 
\begin{equation}
\lambda=2\sqrt{\bigg(\frac{k}{2}(f^{2}-1)\bigg)^{2}+\bigg(\frac{{a}'}{a}\bigg)^{2}}.
\label{lyapunov index1}
\end{equation}
From the comparison between the Lyapunov index \eqref{lyapunov index1} and Lanczos coefficient \eqref{Lancozs coefficient}, these two essential parameters all depict the chaotic features of a dynamical system. Ref. \cite{Bhattacharya:2022gbz,Dymarsky:2021bjq} stated that Lanczos coefficients show the fastest growth which means that $b_n\propto n$. In single-field inflation, we only need to consider the perturbation of the vacuum which will form the large-scale structure of the present universe. By the consideration of Krylov complexity, the impact of excited states will be more essential, represented by Lyapunov index \eqref{lyapunov index1} and Lanczos coefficient \eqref{Lancozs coefficient}. 

Before investigating the Krylov complexity within the method of closed system, we simply analyze the features of Lyapunov index \eqref{lyapunov index1} and Lanczos coefficient \eqref{Lancozs coefficient}. Explicit observations indicate that it will recover the standard dispersion relation as $f=1$. Through a simple algebra, one can obtain $\lambda=\frac{2}{\eta}$, $b_n=\frac{n}{\eta}$ and $K=\frac{1}{4k^2\eta^2}$ ($\eta$ is the conformal time) which is also consistent with \cite{Adhikari:2022oxr}. At the very beginning of the universe with $\eta$ approaches $-\infty$, one can explicitly obtain $K=0$. Especially, one could see that Krylov complexity is very tiny as $k\eta\ll 1$. For the late universe,  the complexity will grow exponentially. These features are perfectly presenting the chaotic features of a dynamical system. However, once taking the impact of modified dispersion relation into account, it clearly shows that the Lyapunov index \eqref{lyapunov index1} and Lanczos coefficient \eqref{Lancozs coefficient} will depend on the momentum $k$ and $f$.  We already know that its growth is linearly proportional to $n$. But, we still need to understand how $k$ and $a$ impact $b_n$ and $\lambda$. 
\begin{figure}
	\centering
	\includegraphics[width=1\linewidth]{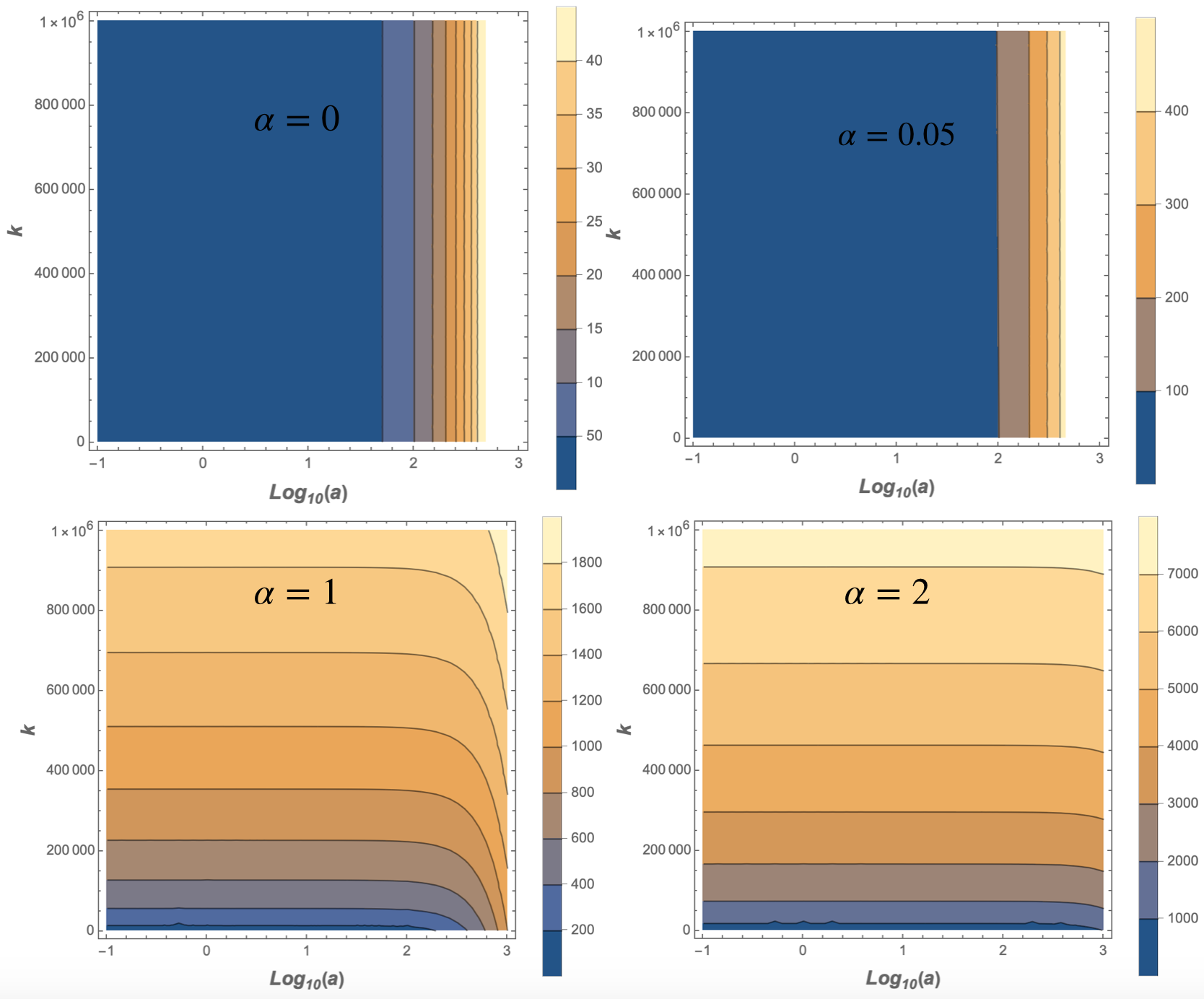}
	\caption{The contourplot of $b_n$ or $\lambda$, where we have set $b_n=\lambda$ with $n=2$ 
		 and $\frac{k_{\rm ph}}{M}=1.5$. We give four various values for $\alpha$ as shown in four cases. The Plot-lengend shows the value of $b_n$ or $\lambda$ as $n=2$. $\log_{10}a$ varies from $-1$ to $3$ and $k$ varies from $1.1\times 10^3$ to $10^6$. }
	\label{fig:bn}
\end{figure}

In Fig. \ref{fig:bn}, we found $b_n=\lambda$ as $n=2$ which is enough to manifest their changing trend. Keeping in mind that $\alpha$ will be larger than zero as $k_{\rm ph}> M$, and $k_{\rm ph}=k/a$. If we set $M=1$ (some certain energy scale), it means that $k$ should be larger than $a$ in Planck units. Meanwhile, $a=10^{y}$, if we set the maximal value of $y=3$, it requires that $k$ must be larger than this value. Through this simple analysis, that is the reason we set the range of $k$ from $1.1\times 10^3$ to $10^6$. The upper panel of Fig. \ref{fig:bn} clearly indicates that $b_n$ and $\lambda$ are nearly determined by $k$, which corresponds to the standard case. By contrast, the modified case shows that $b_n$ and $\lambda$ are mainly determined by $a$. The careful reader may notice that Fig. \ref{fig:bn} does not include the case of $\alpha=3$ which we have confirmed that its behavior is almost the same with $\alpha=2$. Further, it gives us a new hint where the geometrical effects of background are essential for Krylov complexity.

Next, we will evaluate the evolution of Krylov complexity in terms of numerical solution of $r_k$. 
\begin{figure}
	\centering
	\includegraphics[width=1\linewidth]{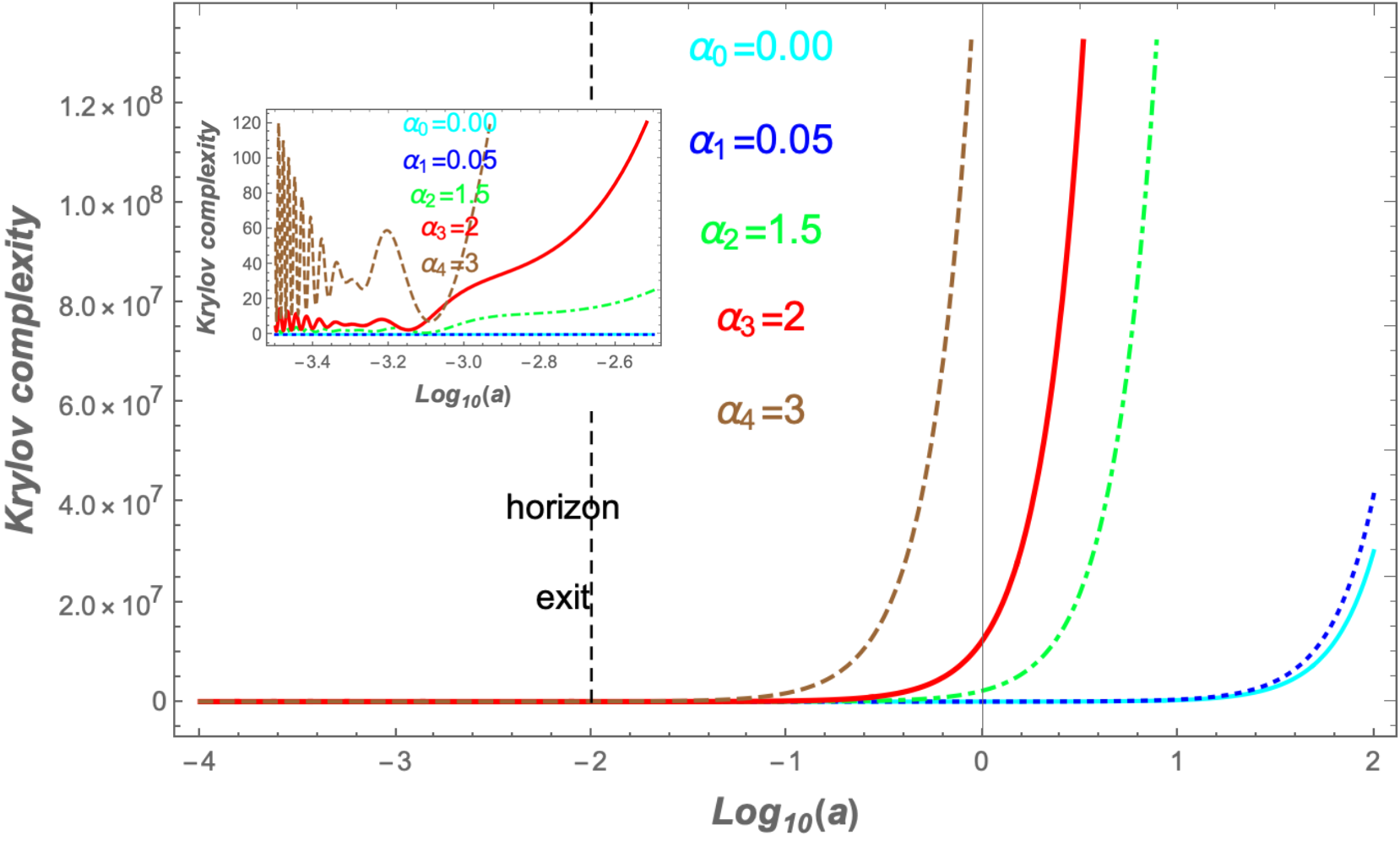}
	\caption{The plot of Krylov complexity \eqref{Krylov complexity1}, $y=\log_{10}(a)$ varies from -4 to 2. The "horizon exit" is located at $y=-2$ (still in the very early universe). We have set $M=H_0=1$.}
	\label{fig:complexity_closed}
\end{figure}
Fig. \ref{fig:complexity_closed} clearly indicates the varying trend of Krylov complexity with various values of $\alpha$ in a method of closed system. One can see that the complexity will grow faster compared with the standard case. To be more precise, the framework of quantum gravity, including the string cosmology, loop gravity $\it e.t.c$, will be more chaotic compared with the standard case. The obvious trend is that the complexity will grow faster as enhancing the value of $\alpha$. For completeness, we also plot the Krylov complexity before the horizon exits, and we find that the Krylov complexity will show the damping oscillation as $\alpha>2$ that is similar to our previous work \cite{Li:2023ekd}. For the standard dispersion relation, one could see that the Krylov complexity is almost zero. Here, we make a conclusion that the complexity of quantum gravity including the Krylov complexity and circuit complexity will lead to the various patterns in the very early universe, whose evolution of circuit complexity can be found in Fig. 4 of \cite{Li:2023ekd}. Generically, this damping oscillation is the criterion for assessing the quantum gravitational models and standard single-field inflationary models. The last thing is that the order of Krylov complexity is much larger than the circuit complexity under similar conditions.

There are other two interesting issues that need to be noticed which have been discussed in Ref. \cite{Adhikari:2022oxr}. One thing is that the Krylov complexity equals the average number of particles in each mode $\left \langle \hat{n}_{k}  \right \rangle =\left \langle \hat{n}_{-k}  \right \rangle=\sinh^{2}r_{k}=K$. The average number density is proportional to the volume, thus one can explicitly obtain that $K\propto~volume$. In some sense, it illustrates the CV conjecture in light of the holographic principle. Another issue is that the corresponding K-entropy can be defined by
\cite{Barbon:2019wsy}, 
\begin{equation}
\begin{split}
S_{K}&=-\sum_{n=0}^{\infty } \left | \phi_{n} \right |^{2}\ln\left | \phi_{n} \right |^{2}=-\sum_{n=0}^{\infty } \frac{\tanh^{2n}r_{k}}{\cosh^{2}r_{k}} \ln\frac{\tanh^{2n}r_{k}}{\cosh^{2}r_{k}}\\&=\cosh^{2}r_{k}\ln(\cosh^{2}r_{k})-\sinh^{2}r_{k}\ln(\sinh^{2}r_{k}),
\end{split}
\label{k entroy}
\end{equation}
where we have made an assumption $\tanh^2(r_k)\textless  1$. If we set $\tanh^2(r_k)=0.8$, the value of $r_k\approx 1.44364$. From Fig. \ref{fig:rk}, one can see the numerics of $r_{k}$, when $r_k=1.44$, we can see that it is almost impossible to find the valid range or $r_k$ as $\alpha \textgreater 1$. And our analytical calculation finds that the K-entropy is divergent when $\tanh^2(r_k)$ is almost one with $r_k\textgreater 3.5$. Also combining with Fig. \ref{fig:rk}, most ranges of $r_k$ will be larger than $3.5$. Here, we cannot give a complete evolution of K-entropy when the scale is the same as Fig. \ref{fig:rk}.

In Ref. \cite{Barbon:2019wsy}, the evolution of Krylov complexity is different where our cases only manifest the growth (nearly exponential growth), in which Ref. \cite{Barbon:2019wsy} shows the Krylov complexity also must saturate after some time scale. One possible reason leading to this is that our analysis is only focused on inflation. If we extend our analysis into the radiation (RD) or the matter domination (MD) period (the epoch after inflation), it may present a similar trend with Ref. \cite{Barbon:2019wsy} left for future work. Another reason is that the method we used is for the closed system.

Let us make a summary of this section, we already use the Lanczos algorithm to calculate the Krylov complexity, Lanczos coefficient, Lyapunov index, and K-entropy. The Hamilton \eqref{standard hamilton} what we implemented is the Hermitian and the wave function of the two-mode squeezed state is unitary. The method what we implement is for the closed system. However, the energy is not conserved for the universe since the temporal symmetry will be broken as the time-reversed. Thus, it is quite natural to extend the analysis into the open system. 

\section{Krylov complexity with approach of open system}
\label{krylov in open}
In this section, we will implement the method of \cite{Bhattacharya:2022gbz} to investigate the Krylov complexity in an open system's method, which the very early universe is dubbed as an open system.

\subsection{Generalized Lanczos algorithm}
 Following  Ref. \cite{Bhattacharya:2022gbz}, the Hamilton includes the dynamic system (closed system), interaction part, and environmental part, where they take $1-d$ transverse-field Ising model as an example whose interaction term is denoted in terms of so-called jump operators (encoded by the Pauli matrices). A general operator in the Heisenberg picture can be written by
  \begin{equation}
  \mathcal{O} (\eta)=e^{i\mathcal{L}_{o}\eta}\mathcal{O},
  \label{general operator} 
  \end{equation} 
where $\mathcal{L}_o$ denotes the Lindbladian presented by the total Hamilton. Using $\mathcal{L}_o$ to act on the Krylov basis, one can get
 \begin{equation}   \mathcal{L}_o|\mathcal{O}_{n})=-ic_{n}|\mathcal{O}_{n})+b_{n+1}|\mathcal{O}_{n+1})+b_{n}|\mathcal{O}_{n-1}),
 \label{generalized recursion relation}
\end{equation} 	
where $c_n$ encodes the information of the open system which is the diagonal part. Compared with the standard relation \eqref{recursion relation}, there is an extra term that is proportional to $c_n$. Its corresponding differential equation for the coefficients are  
\begin{equation}
\begin{split}
{\phi}'_{\eta}=b_{n}\phi_{n-1}(\eta)-b_{n+1}\phi_{n+1}(\eta)+ic_{n}\phi_{n}(\eta)=b_{n}\phi_{n-1}(\eta)-b_{n+1}\phi_{n+1}(\eta)-\tilde{c} _{n}\phi_{n}(\eta)
\end{split}
\label{differential equaiton of b}
\end{equation}
where we have defined $\tilde{c}_n=-ic_n$. Therefore, the expanding coefficients $b_n$ and $c_n$ are explicitly related to the specific Hamilton. 

Recalling that our Hamiton \eqref{standard hamilton} is Hermitian. For the closed part \eqref{closed hamilton}, it only contains the two terms which are proportional to $\hat{c}_{\vec k}^\dagger \hat{c}_{-\vec k}^\dagger$ and $\hat{c}_{\vec{k}} \hat{c}_{-\vec k}$, respectively. For us, the rest two terms which are proportional to $\hat{c}_{-\vec k}^\dagger\hat{c}_{-\vec k}$ and $\hat{c}_{\vec k}\hat{c}_{\vec k}^\dagger$, they are responsible for producing $c_n$ which will be confirmed in the latter investigation. As \cite{Socorro:2022aoz} discussed, the k-essence gravitational system is non-Hermitetian, it strongly indicates the Hamilton of inflationary model is n\label{key}on-Hermitian. However, our studying Hamilton \eqref{standard hamilton} is Hermitian, the problem is that the action \eqref{standard action} is only the quadratic perturbation of Mukhanov variable $v=z\mathcal{R}$. The higher order perturbation of $v$ could give rise to the non-Hermitian terms,  such as $v^3$, $\partial_iv v^2$ $\it e.t.c,$ that could give the non-Hermitian terms of Hamilton.  Meanwhile, we also neglect the potential of inflation describing the interaction between the dynamic system and the enviroment. Here we give more details for readers who are not familiar with the perturbation theory of inflation. In light of \cite{Baumann:2009ds}, we 
have the quadratic action in terms of the Mukhanov variable as shown
\begin{equation}
S=\frac{1}{2}\int d\eta dx^3\bigg[ (\partial_\eta v)^2+k^2v^2+\big(\frac{a''}{a}-\frac{d^2V}{d\phi^2}\big)v^2 \bigg]
\end{equation}
 in momentum space where $\phi$ is the inflaton field (background field). If the potential has the Yukawa coupling term $\lambda_s \phi ^2\chi^2$ and we keep it in the action that also will lead to the Non-Hermitian terms. Due to the slow-roll conditions $\frac{a''}{a}\gg \frac{d^2V}{d\phi^2}$, one could neglect the contribution of potential.   
 
  From another aspect, once the environment begins to affect, the non-Hermitian term will make the growth behavior of the Lanczos coefficient in the integrable region and the chaotic region indistinguishable, so that the open system and the closed system are almost indistinguishable under the Hermitian recursive algorithm \cite{Zanardi:2020hsu,Andreadakis:2022hqt}. Keeping in mind that the Hermitian recursive algorithm is the Hamilton \eqref{standard hamilton} in our paper. Although Hamilton is Hermitian, the wave function of the dynamical system is not a two-mode squeezed state anymore.

\subsection{General discussions of $\phi(n)$}
\label{general discussions}
After analysis, we will proceed with the investigation for the asymptotic quantitative behavior of $\phi_n$. Recursive relation \eqref{differential equaiton of b} is the discrete version of the partial differential equation (PDE) of $\phi_n$, then we will transit it into the continuous version of PDE for $\phi(n)=\phi_n$ (a smooth function with $n$) as follows, 
\begin{equation}
\partial _{\eta}\phi +2b_n\partial_n\phi +\tilde{c}_n\phi=0,
\label{eom of phin0}
\end{equation}
where we have made the approximations: $b_{n+1}\approx b_n$ and $\phi_{n+1}-\phi_{n-1}\approx2\partial_n \phi$. Following Ref. \cite{Bhattacharjee:2022lzy}, its contious version can be written by $\partial _{\eta}\phi +n[\chi\mu\phi+2\alpha\partial_{n}\phi ]=0$. Making a comparison, one can obtain $\tilde{c}_n=n\chi\mu$ and $b_n=n\alpha$. As for $\eta$ approaches to infinity, Eq. \eqref{eom of phin0} will be of the form $\phi(n)\propto e^{-n/\xi}$ with $\xi=\frac{2\alpha}{\chi\mu}$. Since $\mu$ presents the strength between the dynamical system and the environment, one can consider that the dynamical system is nearly a closed system as $\mu$ approaches zero and vice versa. Accordingly, $\xi$ will be very large as $\mu$ is very tiny. The corresponding evolution of complexity in the weak dissipation region will be in this way: At the very beginning of dynamical system, the Krylov complexity will spread exponentially whose average position is $K(\eta)\propto e^{2\alpha \eta}$ as $\eta<\eta_*$, which it is still true until $K(\eta)$ saturates to $\xi$ and $\eta>\eta_*$, where $\eta_*\sim \frac{1}{2\alpha}\ln (\frac{2\alpha}{\chi \mu})$. It is straightforward to see that the $\eta_*$ is just the $\xi$ nothing else. As for the strong dissipation region, the growth of Krylov complexity is different.

Finally, we make a comment on this section. One can see that the analysis of PDE \eqref{eom of phin0} is very generic. Since we only implement the generalized recursive relation \eqref{generalized recursion relation} and Schr$\ddot{o}$dinger to derive the PDE of $\phi(n)$, which is Schr$\ddot{o}$dinger equation essentially. The information of various models is encoded in $c_n$ and $b_n$. We also need to emphasize that the above analysis is only valid in the weak dissipative region. And $\eta$ is just a parameter that is not encoding the information of conformal time in this section.

\subsection{Wave function of $\phi(n)$ in a weak dissipative system}
\label{wave function in oepn system}
Recalling that we can decompose the Hamilton \eqref{standard hamilton} into closed part and open part as follows, 
\begin{equation}
\mathcal{L}_o=\mathcal{H}_o=\mathcal{H}_{closed}+\mathcal{H}_{open}, 
\label{decompsition of H}
\end{equation}
where we have found that $\mathcal{H}_{closed}= \bigg[\frac{k}{2}(f^{2}-1)+i\frac{a{}'}{a}\bigg]\hat{c}_{k}^{\dagger }\hat{c}_{-k}^{\dagger }+\bigg[\frac{k}{2}(f^{2}-1)-i\frac{a{}'}{a}\bigg]\hat{c_{k}}\hat{c}_{-k}$, as for $\mathcal{H}_{open}$, it naturally reads as 
\begin{equation}
	\mathcal{H}_{open}=\frac{k}{2}(f^{2}+1)(\hat{c }_{-k}^{\dagger}\hat{c}_{-k}+\hat{c_{k}}\hat{c}_{k}^{\dagger }).
	\label{open hamilton}
\end{equation}
Using Hermitian recursive relation to act on $\mathcal{O}_{n}$, one can straightforwardly obtain as follows
\begin{equation}
\begin{split}
\mathcal{L}_{o}|\mathcal{O}_{n})=(2n+1)\frac{k}{2}(f^{2}+1)|\mathcal{O}_{n})+(n+1)(f^{2}-1)+i\frac{z{}'}{z})|\mathcal{O}_{n+1})+n(\frac{k}{2}(f^{2}-1)-i\frac{z{}'}{z})|\mathcal{O}_{n-1}). 
\end{split}
\label{recursive relation in open system}
\end{equation}
Compared with \eqref{generalized recursion relation}, one can explicitly obtain 
\begin{equation}
c_{n}=i(2n+1)\frac{k}{2}(f^{2}+1),
\label{cn}
\end{equation}
where $b_n$ is \eqref{Lancozs coefficient}. Being armed with these quantities, one can rewrite Eq. \eqref{eom of phin0} into 
\begin{equation}
\partial _{\eta}\phi +n\bigg[\bigg(2+\frac{1}{n}\bigg )\frac{k}{2}\bigg(1+f^{2}\bigg)\phi+2b_{n}\partial_{n}\phi \bigg]=0.
\label{eom of phin2}
\end{equation}
With the definition of $c_n=i\mu\chi n$ and $b_n=\alpha n$, we can get the following relation as follows
\begin{equation}
\chi\mu=\bigg(2+\frac{1}{n} \bigg)\frac{k}{2}\bigg(1+f^{2}\bigg).
\label{chimu}
\end{equation}
We do a simple analysis for $\chi\mu$. Here, $\chi$ is a free parameter. In the later investigation, we will show that $\chi\mu$ will play a role in the dissipative coefficient. For completely solving PDE \eqref{eom of phin2}, we will make a Fourier transformation between $\eta$ and frequency $\omega$, 
 \begin{equation}
 \phi(\eta)=\int_{-\infty}^{+\infty} f(\omega)e^{i\omega\eta}d\omega,
 \label{fourier transformation}
 \end{equation} 
 \begin{equation}
 i\omega f(\omega)+n\chi\mu f(\omega)+n2\alpha\partial_{n}f(\omega)=0,
 \label{eom of phin3}
 \end{equation}
 its corresponding solution can be derived explicitly as follows,
  \begin{equation}
  f(\omega)=Cn^{-\frac{in\omega}{2b_{{n}}}}e^{-\frac{n^{2}\chi\mu}{2b_{n}}}=Cn^{-\frac{in\omega}{2b_{{n}}}}e^{-\frac{n}{\xi}},
  \label{phin3}
  \end{equation}
where $C$ is a coefficient with 
  \begin{equation}
  \xi=\frac{2\sqrt{(k/2)^2(f^2-1)^2+(a'/a)^2}}{(2+1/n)(k/2)(1+f^2)}.
  \label{xi}
  \end{equation}
and we have obtained the known factor $e^{-\frac{n}{\xi}}$. Until present, we have obtained the solution without any assumption or approximation. Recalling that $\eta$ is just a parameter that does not represent the conformal time since we obtain it only acoording to \eqref{phin3}. A careful reader could notice that $\eta$ cannot be larger than zero in inflation in light of $\eta=\frac{-1}{aH}$
where the range of $\eta$ is $-\infty<\eta<0$. It seems that is conflicted with $\eta$ approaches infinity. The truth is that we only consider inflation where the formula of the scale factor is determined. If it is extending our analysis into the late universe, the value of $\eta$ can be the infinity, such as $\eta\propto t^\alpha$ ($t$ is the physical time and $\alpha>0$). Consequently, one can conclude that the inflationary period is a strong dissipative system. If one expects to analyze the weak dissipative region, we must extend into the late universe. Thus, the construction of an exact wave function for describing inflation is quite necessary. The function of weak dissipative approximation is that we could test the validity of our exact wave function. Since all of the analysis should be recovered into the case of a closed system under this approximation. In the next section, we will construct an exact wave function for describing a strong dissipative system.

\subsection{Exact wave function of $\phi(n)$ in an open inflationary universe}
\label{exact wave function}
Before constructing the exact wave function of $\phi_n$, we will present the validity of the method coming via \cite{Parker:2018yvk}, in which they stated that their whole analysis is in light of the asymptotic behavior of $b_n^{(1)}$ (Lanczos coefficient), 
\begin{equation}
	b_n^{(1)}=\alpha n+\gamma,
	\label{bn0}
\end{equation}
where $\alpha$ and $\gamma$ are real constants determined by various models. This kind of behavior of $b_n$ is quite universal for an infinite, non-integrable, many-body system. In our case, the $b_n$ \eqref{Lancozs coefficient} is of the linear growth depending on $n$ which is perfectly agreed with their assumption. Although this kind of method is for the closed system, one can naturally promote it into the open system \cite{Bhattacharjee:2022lzy} due to the high symmetry of $\mathcal{L}_o$ corresponding to the total Hamilton \eqref{standard hamilton}. Based on these two points, we will use the generalized way of constructing the exact wave function. First, we will define the two parameters as follows:
 \begin{equation}
b_n^2=|1-u_1^2|n(n-1+\beta)~~~~~~~~~c_n=iu_2(2n+\beta),
 \label{bn and cn}
 \end{equation}
 where $b_n$ and $c_n$ (corresponding to $a_n$ of \cite{Bhattacharjee:2022lzy}) are valid for many specific models, only requiring that $b_n \propto n$, and $u_1$ and $u_2$ are encoding the information of various models.  
 Comparing with \eqref{Lancozs coefficient} and \eqref{cn}, one can straightforwardly obtain
 \begin{equation}
 	\beta=1,~~~~|1-u_1^2|=\frac{k^2}{4}(f^2-1)^2+\frac{a'^2}{a^2},~~~~~u_2=\frac{k}{2}(f^2+1). 
 	\label{u1 and u2}
 \end{equation}
 By improving the original version, we define two parameters $u_1$ and $u_2$. In the original version, they only consider $u$ corresponding to $u_1=u_2$ in our case, where they investigate the Krylov complexity by considering $u$ as some constant values. In our case, the situation is different since the structure of $b_n$ and $c_n$ are exactly determined by our Hamilton \eqref{standard hamilton}, which also precisely determines the formula of $u_1$ and $u_2$ as showing Eq. \eqref{u1 and u2}. Ref. \cite{Bhattacharjee:2022lzy} only studied the weak dissipative region, in which the Krylov complexity will exponentially grow and then approach some constant values. But, the practical situation will be more complicated, such as $u_1$ and $u_2$ in our case. In the later investigation, we will proceed with the situation when $u_1$ and $u_2$ are larger than unity, which would lead to the different evolution patterns of Krylov complexity. Being armed with these two parameters and following the detailed calculation of \cite{Parker:2018yvk},
 one can explicitly obtain the exact wave function as follows:
 \begin{equation}
 \phi_n(\eta)=\frac{\rm sech(\eta)}{1+u_2\rm{tanh}(\eta)}\times |1-u_1^2|^{\frac{n}{2}}\bigg(\frac{\tanh(\eta)}{1+u_2\tanh(\eta)}\bigg)^n.
 \label{exact wave function1}
 \end{equation}
 This wave function is our central result in this paper which could desribe the strong and weak dissipative system, of course including the inflation.

 First, we need to check that the exact wave function can restore into the case of a closed system under the weak dissipative system, but it is after inflation. As $\eta$ approaches infinity,  the momentum will decrease dramatically due to the exponential expansion of the universe, thus one can see that $u_1$ and $u_2$ will be much smaller than unity. Another way is that the wave function \eqref{exact wave function1} will behave like 
 \begin{equation}
 	\phi_n(\eta\rightarrow \infty)\propto \bigg(\frac{\sqrt{|1-u_1^2|}}{1+u_2}\bigg)^2,
 \end{equation}
 where we have set $n=1$. Doing the Taylor expansion of $u_1^2$ and $u_2$ of wave function \eqref{exact wave function1}, one can obtain $\phi_n(\eta\rightarrow \infty)\propto 1-u_2-\frac{1}{2}u_1^2$. According to the general discussion of \ref{general discussions}, the wave function of $\phi_n$ should have the form as $\exp(-\frac{n}{\xi})$, its Taylor expansion for the firs order is $1-\frac{1}{\xi}=1-u_2-\frac{1}{2}u_1^2u_2$. As $u_1\ll 1$ and $u_2\ll 1$, thus we could see that the asymptotic wave function could match the stationary solution $\phi_n\propto e^{-n/\xi}$. Thus, one can also derive that the $\xi^{-1}\approx u_2$ in the weak dispassion region so as to obtain $\eta_*=\ln(1/u_2)/2$. Until present, our discussion is valid for the weak dissipative system.

 Next, we will calculate the Krylov complexity in light of its definition $K=\Sigma_n n |\phi_n|^2$, which can be explicitly obtained by 
 \begin{equation}
 	K=\frac{\rm{sech^2\eta}|1-u_1^2|\tanh^2\eta}{\big[1+2u_2\tanh\eta+(u_2^2-|1-u_1^2|)\tanh^2\eta\big]^2},
 	\label{krylov complexity in open}
 \end{equation}
 where we also implemented the identity $\sum_{n=1}^{+\infty}n z^n=\frac{z}{(1-z)^2}$. Following the assumption with $u_1,~u_2\ll 1$, meanwhile, we only retain the first order of $u_1$ or $u_2$. Consequently, we could ignore the contribution of $u_1^2+u_2^2$ and expand $u_2$ around zero, one could obtain the following asymptotic formula
 \begin{equation}
   K=\sinh^2\eta-4u_2\sinh^3\eta\cosh\eta +\mathcal{O}^{n}(u_2),
   \label{asympotic formula of K}
 \end{equation}
it clearly shows that the Krylov complexity will be recovered into the form of $\sinh^2\eta$, where the Krylov complexity is $\sinh^2 (r_k)$ in a closed system whose wave function \eqref{operator wave function} comes via the two-mode squeezed state. The higher order will contribute to the correction due to the dissipation effects. However, all of these calculations hold true under the assumption with $u_1,~u_2\ll 1$ which furtherly tests the validity of the exact wave function \eqref{exact wave function1}. Following the previous discussion, the inflation is a strong dissipative system, we will show the numerics of Krylov complexity \eqref{asympotic formula of K}.

\subsection{Exact evolution of Krylov complexity in an open inflationary universe}
 Once obtaining the exact formula of Krylov complexity, one can precisely investigate the evolution of Krylov complexity. We have analyzed that the values of $u_1$ and $u_2$ will be larger than one. Observing that the values of $u_1$ and $u_2$ are mainly determined by $k$ and $a$. Recalling that the universe undergoes an exponential expansion during an inflationary period, the small values of $k$ could represent the evolution of Krylov complexity as inflation almost finishes. At the very beginning of inflation, the values of $k$ should be large. Consequently, one could see a complete evolution of Krylov complexity when $k$ varies from tiny values to large values. 
 
 \begin{figure}
 	\centering
 	\includegraphics[width=1\linewidth]{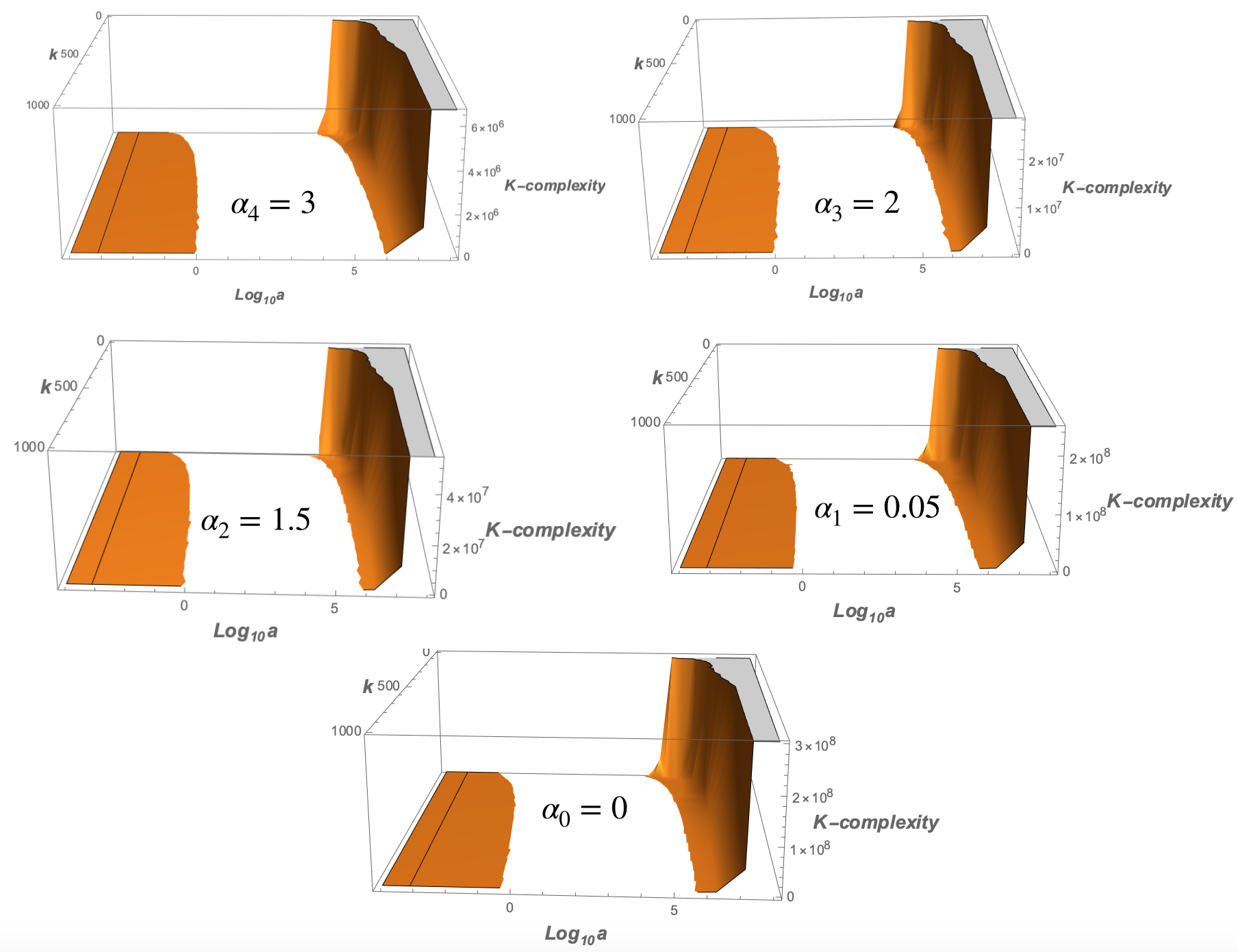}
 	\caption {The precise evolution of Krylov complexity in terms of \eqref{krylov complexity in open}. The range of $k$ is from $0$ to $1000$. The range of $\log_{10}(a)$ is from $-4$ to $8$. We have set $M=H_0=1$.}
 	\label{fig:complexity_open}
 \end{figure}
\begin{figure}
  	\centering
  	\includegraphics[width=1\linewidth]{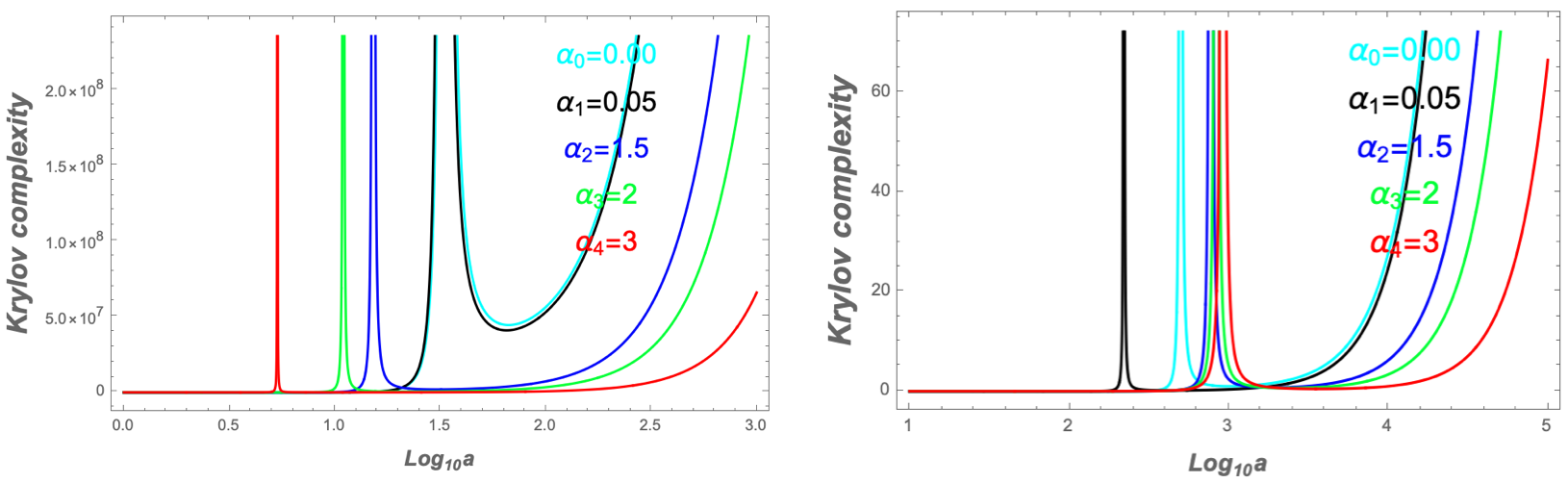}
  	\caption {The left pannel corresponds to $k=0.01$ as $0<y<3$. The right pannel matches $k=1000$ as $1<y<5$. We have set $M=H_0=1$.}
  	\label{fig:complexity_open1}
  \end{figure}

 First, we will give the precise evolution of Krylov complexity in terms of \eqref{krylov complexity in open} represented by Fig. \ref{fig:complexity_open}. One can see that the total trend of Krylov complexity is almost the same with various values of $\alpha$, the only difference is the order of the Krylov complexity, the standard dispersion relation will hold the largest Krylov complexity under the same parameters, which is contradicted with the evolution in closed system presenting in Fig. \ref{fig:complexity_closed}, where the modified case will grow faster compared with the standard case. Observing that there is a blank area for every case, which means that the Krylov complexity will decrease to zero and then it will experience exponential growth. One can also find that the varying trend of Krylov complexity is mainly determined by scale factor $a$, not $k$. In light of this observation, we could fix some specific values of $k$ and vary with respect to $y$ in order to check the evolution trend of Krylov complexity. For completeness, we only consider two cases: (a). $k=0.01$; (b). $k=1000$, which could provide a full perspective of the evolution trend. Fig. \ref{fig:complexity_open1} clearly indicates that the Krylov complexity will reach the peak of various $\alpha$, then it will dramatically decrease to zero. After some time scale, it will show the exponential growth pattern, in which the standard case will grow faster compared with modified cases. Thus, Fig. \ref{fig:complexity_open1} will confirm the total varying trend as shown in Fig. \ref{fig:complexity_closed}. In some sense, Fig. \ref{fig:complexity_open1} has given a precise and complete evolution of Krylov complexity.

 Although both the modified and standard cases show an exponential growth trend throughout the inflationary period, it is worth noting that the modified case in the method of open system will grow at a slower pace than the standard case. But the Krylov complexity of the modified case grows faster than that of the standard case in the closed system's method, as shown in Fig. \ref{fig:complexity_closed}. We need to identify the factors responsible for this difference. The evolution of Fig. \ref{fig:complexity_open1} comes from the precise formula of Krylov complexity \eqref{krylov complexity in open}. In comparison to the Krylov complexity \eqref{Krylov complexity1} in the closed system method, it is explicit that the Krylov complexity in open system's method is influenced by the dissipation effects encoding in dissipative coefficient $u_2$. Mathematically speaking, one can easily observe the value of $u_2$ \eqref{u1 and u2} is determined by $f^2$. If the value of $f^2$ is enhanced, the value of Krylov complexity will be suppressed as shown in Eq. \eqref{krylov complexity in open}. Therefore, it is easy to conclude that the strong dissipative effects are responsible for the difference in Krylov complexity's evolution between the open system method and the closed system method.

Another big issue is that the Krylov complexity will not reach a constant after some certain scale ($\eta_*$ in Sec. \ref{general discussions}), which is not consistent with Ref. \cite{Bhattacharjee:2022lzy}. Here, we give two possible reasons that lead to this phenomenon: (a) The lasting time of inflation is too short that is from $10^{-36}$ sec to $10^{-32}$ sec. The Krylov complexity may show a similar trend with \cite{Bhattacharjee:2022lzy} by extending our investigations into RD or MD, where the formula of $a$ is different in MD or RD. (b) As Fig. \ref{fig:bn} shows, the $b_n$ and $\lambda$ will be mainly determined by the scale factor $a$. Thus, it is reasonably assumed that the exponential expansion will cause the continuous growth of Krylov complexity from the large scale, especially for the modified dispersion relations. From the small scale, we can see that there is a peak of complexity, this peak appears after the horizon exits no matter whether the initial value of $k$ is small or huge. According to the previous discussion, the transition from the quantum level to the classical level may lead to the occurrence of this peak whose nature is still mysterious, especially for the decoherence process. It should be noted that the decoherence will cause the circuit complexity to saturate to some constant values \cite{Bhattacharyya:2022rhm},  it may be applied to the Krylov complexity. Thus, one could see that the decoherence may play a vital role in the evolution of Krylov complexity in inflation. 

Finally, we should compare the Krylov complexity in methods of closed system and open system. The reason we need to investigate the Krylov complexity in a method of open system is that the whole universe is an open system.  Ref. \cite{Cheung:2007st} has shown the universe will break the symmetry along the temporal part, which leads to the non-conservation of energy for the universe. From another perspective, the generation of particle is a non-equilibrium process encoding in the preheating period \cite{Kofman:1994rk,Kofman:1997yn}, which is an open system. Based on these two factors, it is quite natural to consider inflation as an open system. Consequently, the method of utilizing the method of an open system for investigating the Krylov complexity is more realistic and reliable compared with the method of closed system.

  \subsection{Krylov entropy in an open inflationary universe}
  \label{krylov entropy}
   
   We have checked that the wave function \eqref{exact wave function1} could nicely recover the case of the weak region, meanwhile it also can restore into the case of a closed system. For completeness, we will also calculate its corresponding Krylov entropy.

 Being armed with the definition of Krylov entropy \eqref{k entroy} and its resulting formula is derived in Appendix \ref{App:a} that reads as follows,  
 \begin{equation}
 \begin{split}
 	S_k&=-\frac{\rm sech^2\eta}{A^2}\bigg[-(1+u_2\tanh\eta)^2\ln(1+u_2\tanh\eta)^2-\tanh^2\eta|1-u_1^2|\ln\rm sech^2\eta
 		\\& +(1+u_2\tanh\eta)^2\ln\rm sech^2\eta+\tanh^2\eta|1-u_1^2|\ln(1-u_1^2)+\tanh^2\eta|1-u_1^2|\ln\tanh^2\eta\bigg], 
 \end{split}
 \label{k entropy in open}
 \end{equation}
 where $A$ is defined in Eq. \eqref{A}. Once obtaining the exact formula of Krylov entropy, we expect our Krylov entropy could recover in the case of a closed system whose formula are determined by \eqref{k entroy}. Following the same assumption of $u_1\ll 1$ and $u_2\ll 1$, thus one can neglect the higher order of $u_1$ and $u_2$, namely $u_1^2$ and $u_2^2$ in our case. Then, Eq, \eqref{k entropy in open} will become
\begin{equation}
\begin{split}
	&S_k\approx-\big(\cosh^2\eta-4u_2\cosh^3\eta\sinh\eta\big)\bigg[-2u_2\tanh\eta+\rm sech^2\eta\ln \rm sech^2\eta+2u_2\tanh\eta\ln\rm sech^2\eta
	\\&+\tanh^2\eta\ln\tanh^2\eta\bigg]
	\\&=-\sinh^2\eta\ln\tanh^2\eta-\ln\rm sech^2\eta+u_2\big(2\cosh\eta\sinh\eta+2\cosh\eta\sinh\eta\ln\rm sech^2\eta
	\\&+4\cosh\eta\sinh^3\eta\ln\tanh^2\eta\big)
	\\&=\cosh^2\eta\ln\cosh^2\eta-\sinh^2\eta\ln\sinh^2\eta+u_2\big(2\cosh\eta\sinh\eta+2\cosh\eta\sinh\eta\ln\rm sech^2\eta
	\\&+4\cosh\eta\sinh^3\eta\ln\tanh^2\eta\big),
\end{split}
\label{krylov entropy2}
\end{equation}
where we have used $-\rm sech^2\eta/A^2=-\cosh^2\eta+4u_2\cosh^3\eta\sinh\eta$ expanding around zero of $u_2$ and we only retain the first order of $u_1$ or $u_2$.  For the leading order, it is consistent with the formula of \eqref{k entroy} which confirms the wave function \eqref{exact wave function1} is correct. The correction that is proportional to $u_2$ comes via the dissipation effects. 

Let us make some comments for the whole section. $(a).$ All of these calculations are based on the precise wave function \eqref{exact wave function1}. This method is model independent whose details can be found in Appendix D of Ref. \cite{Parker:2018yvk}, only requiring that $b_n\propto n$. Consequently, the Krylov complexity \eqref{krylov complexity in open} and Krylov entropy \eqref{k entropy in open} is applicable to the infinite, many-body, chaotic system. $(b).$ Due to the inflation, it will dramatically change the evolution of Krylov complexity in the inflationary period, in which the whole trend will grow and will not saturate to some values. $(c).$ Since the perturbation of curvature will transit from quantum level to classical level, which may lead to the peak of complexity at a small scale, and it even will cause the complexity will saturate some values by considering the Non-Hermitian terms in our case \cite{Bhattacharyya:2022rhm}. Finally, one could discover that the method for calculating the Krylov entropy in an open system will be more convenient compared with the circuit complexity, where we only need the wave function to calculate its Krylov complexity and Krylov entropy. In a traditional method, we need the numerics of $\phi_k$ and $r_k$ to simulate the corresponding circuit complexity, where their numerical solutions are highly dependent on the initial conditions.

\section{conclustions and outlook}
\label{conclusion}
Krylov complexity could provide a unique definition for calculating the complexity, which is free of ambiguities compared with the traditional computational complexity. With the application of quantum information theory, the Krylov complexity will be more and more important in high-energy physics. From another aspect, inflation naturally gives a platform for high energy physics, such as the generation of particles, $\it e.t.c$, where the most important place is that the background of the universe will experience an exponential expansion within a very short temporal interval that may lead to the difference of Krylov complexity. The original method for calculating the Krylov complexity is based on static and flat spacetime. How to develop the method for calculating the Krylov complexity in inflation is of particular interest. In this paper, we focus on the framework of modified dispersion relation in inflation. Since many quantum gravitational frameworks could lead to this kind of modified dispersion relation, our analysis is naturally applicable to them, such as the string cosmology, loop gravity, Lorentz violation, and so on. 

Due to the extremal importance of inflation in high energy physics and cosmology, we systematically calculate the Krylov complexity in closed system's method and open system's method, where the system is the universe itself. Here, we summarize the obtained results. 

$(a).$ Sec. \ref{Krylov complexity} gives a brief introduction to Krylov complexity and calculates the Krylov complexity and Lanczos coefficient in the closed system's method. Sec. \ref{modified dispersion relation and inflation} simply introduces the modified dispersion relation in inflation, then we also give some very fundamental information about inflation whose more details can be found in \cite{Baumann:2009ds}. Based on these two concepts, we found that $b_n$ is proportional to $n$ as shown in Eq. \eqref{Lancozs coefficient}, thus it could be seen that our universe is an infinite, many-body, and chaotic system confirmed by \cite{Bhattacharya:2022gbz}. Meanwhile, Fig. \ref{fig:bn} indicates that the standard dispersion case of the Lanczos coefficient and Lyapunov index are mainly determined by scale factor (single field inflation), the modified case is mainly determined by momentum $k$, where these two parameters present the chaotic features of the system. To be more precise, the scale factor could mainly determine the growth of the operator and chaos for the standard case. As for the modified case, it was mostly influenced by the momentum.

$(b).$ Fig. \ref{fig:complexity_closed} clearly indicates that the Krylov complexity will show the irregular oscillation before the horizon exits which shows a similar pattern compared with circuit complexity \cite{Li:2023ekd}. And the modified case will lead to the faster growth of complexity (exponential growth). We also found that the number density is the same with Krylov complexity similar to \cite{Adhikari:2022oxr}. Since the number density is proportional to volume, we illustrate the CV conjecture in some sense. Ref. \cite{Barbon:2019wsy} shows there is a saturation value of Krylov complexity. In our case, it may show a similar trend of complexity by extending our analysis into RD and MD. Meanwhile, the method what we consider is for a closed system. 

$(c).$  Refs. \cite{Cheung:2007st,Kofman:1994rk,Kofman:1997yn} has indicated that our universe is an open sytem. Thus, the method of open system treating the Krylov complexity is more realistic and reliable compared with that of closed system. In the approach of an open system, the investigation is based on our central result, an exact wave function \eqref{exact wave function1} describing the strong and weak dissipative system, including inflation. Compared with the original method for constructing the wave function \cite{Bhattacharjee:2022lzy}, our wave function \eqref{exact wave function1} could describe the strong and weak dissipative region. According to $\eta=-\frac{1}{aH}$, one can see that $\eta$ cannot be larger than zero in inflation. But the situation may be different after inflation since $\eta\propto t^\alpha$ in the late universe. Thus, one can conclude that the inflation is a strong dissipative system. However, the analysis can be dubbed as a criterion for testing the validity of the exact wave function \eqref{exact wave function1}. As a further test, the calculation for Krylov complex \eqref{asympotic formula of K} and Krylov entropy \eqref{krylov entropy2} could nicely recover into the case of a closed system. In the weak dissipative region, our calculation could return into the analysis of \cite{Bhattacharjee:2022lzy}, where there is scrambling time dubbed as a boundary for studying the complexity. The scrambling time is just $\xi$ \eqref{xi} which is a variable determined by $k$ and $a$. For the strong dissipative region, one could see that the varying trend is always growing from the large scales, which will not reach some constant values. Fig. \ref{fig:complexity_open} clearly indicates the evolution trend of Krylov complexity. Generally, one cannot distinguish the standard case and modified case from the strong dissipative region. At the small scale, there is also one peak of Krylov complexity. One can see that inflation will have a huge impact on the Krylov complexity.

Our work is just the preliminary check of Krylov complexity in cosmology. There are still lots of ideas that need to be investigated in the future. Here, we list several problems for future research. 

Firstly, our numerics has shown that varying trend of Krylov complexity is enhancing at the whole inflation, which is not consistent with \cite{Bhattacharjee:2022lzy}, where they show that Krylov complexity will saturate to some constant values. How to confront their results, one possible solution is extending our analysis into the MD or RD. Since the scale factor will evolve differently in these two periods. The various scale factors may cause the Krylov complexity to be saturated to some constant values. Another possibility comes via the decoherence of the curvature perturbations. Ref. \cite{Bhattacharjee:2022lzy} has investigated the decoherence of the circuit complexity will saturate the constant values, which may be valid for our case.

Secondly, all of our analysis is based on the Hamilton \eqref{standard hamilton} that is Hermitian. The original method for evaluating the Krylov complexity in an open system is based on the Lindblad master equation \cite{Lindblad:1975ef,Gorini:1975nb}. The corresponding Lindblad of various dynamical systems is non-Hermitian. For obtaining the non-Hermitian Hamilton, one could expand the action into the higher order in terms of the Mukhanov variable. Another way is that we could add the potential of inflaton in MD or RD since the contribution of potential cannot be ignored due to the violation of slow-roll condition. 

Thirdly, the horizon exits will lead to the curvature perturbation transit from the quantum level to the classical level. Thus, this process may impact the evolution of Krylov complexity. Especially, one can expand the action of inflaton to the higher order to produce the minimal decoherence \cite{Burgess:2022nwu}, which only incorporates the self-interaction via General Relativity. We are still curious about whether the holographic principle can be applied to this inflationary period or not, just like the CV or CA conjecture.

 \section*{Acknowledgements}
 LH appreciates the fruitful discussions with Ai-Chen Li about the Hermitian properties of Hamilton. We are also grateful for the critical reading of this manuscript from Hai-Qing Zhang, Xin-Fei Li, Zhuo-Ran Huang, and Ge Gao. LH and TL are funded by NSFC grant NO. 12165009 , Hunan Natural Science Foundation NO. 2023JJ30487 and NO. 2022JJ40340. 

\appendix
\section{The calculation of Krylov entropy}
\label{App:a}
In this appendix, we will proceed with the calculation of Krylov entropy step by step. First, we will write down its explicit formula
\begin{equation}
\begin{split}
	S_k=&-\sum_{n=0}^{+\infty}|\phi_n|^2\ln|\phi_n|^2=-\sum_{n=0}^{+\infty}|\phi_n|^2\bigg[\ln\frac{\rm sech^2\eta}{(1+u_2\tanh\eta)^2}+n\ln|1-u_1^2|
	\\&+n\ln\frac{\tanh^2\eta}{(1+u_2\tanh\eta)^2}\bigg],
\end{split}
\label{k total}
\end{equation}
where $\phi_n$ is the exact wave function \eqref{exact wave function1}. 
Then, we will decompose it into three parts, respectively. 
\begin{equation}
\begin{split}
&S_{k_1}=-\sum_{n=0}^{+\infty}|\phi_n|^2\ln\frac{\rm sech^2\eta}{(1+u_2\tanh\eta)^2},
\\&S_{k_2}=-\sum_{n=0}^{+\infty}n|\phi_n|^2\ln|1-u_1^2|,
\\&S_{k_3}=-\sum_{n=0}^{+\infty}n|\phi_n|^2\ln\frac{\tanh^2\eta}{(1+u_2\tanh\eta)^2}
\end{split}
\label{k123}
\end{equation}
We will calculate $S_{k_1}$ first,
\begin{equation}
\begin{split}
	&S_{k_1}=-\sum_{n=0}^{+\infty}|\phi_n|^2\ln\frac{\rm sech^2\eta}{(1+u_2\tanh\eta)^2}\\
	&=-\frac{\rm sech^2\eta}{(1+u_2\tanh\eta)^2}\ln \frac{\rm sech^2\eta}{(1+u_2\tanh\eta)^2}\sum_{n=0}^{+\infty}\Bigg[\frac{|1-u_1^2|\tanh^2\eta}{(1+u_2\tanh\eta)^2}\Bigg]^n\\
	&=-\frac{\rm sech^2\eta}{(1+u_2\tanh\eta)^2}\ln \frac{\rm sech^2\eta}{(1+u_2\tanh\eta)^2}\frac{(1+u_2\tanh\eta)^2}{1+2u_2\tanh\eta+(u_2^2-|1-u_1^2|)\tanh\eta}\\
	&=-\frac{\rm sech^2\eta}{1+2u_2\tanh\eta+(u_2^2-|1-u_1^2|)\tanh\eta}\ln\frac{\rm sech^2\eta}{(1+u_2\tanh^2\eta)^2}
\end{split}
\label{k1}
\end{equation}
As for $S_{k_2}$, we can also do the following calculation,
\begin{equation}
	\begin{split}
	S_{k_2}&= -\sum_{n=0}^{+\infty}n|\phi_n|^2\ln|1-u_1^2|\\
	&=-\frac{\rm sech^2\eta\ln|1-u_1^2|}{(1+u_2\tanh\eta)^2}\sum_{n=0}^{+\infty}n\bigg[\frac{|1-u_1^2|\tanh^2\eta}{(1+u_2\tanh\eta)^2}\bigg]^n\\
	&=-\frac{\rm sech^2\eta\ln|1-u_1^2|}{(1+u_2\tanh\eta)^2}\frac{(|1-u_1^2|\tanh^2\eta)/(1+u_2\tanh\eta)^2}{\big[(1-(|1-u_1^2|\tanh^2\eta)/(1+u_2\tanh\eta)^2\big]^2}\\
	&=-\frac{\rm sech^2\eta\tanh^2\eta|1-u_1^2|\ln|1-u_1^2|}{(1+2u_2\tanh\eta+(u_2^2-|1-u_1^2|)\tanh\eta)^2}\ln\frac{\rm tanh^2\eta}{(1+u_2\tanh^2\eta)^2}
	\end{split}
	\label{k2}
\end{equation}
Similarly with $S_{k_2}$, we can obtain $S_{k_3}$ as follows,
\begin{equation}
	\begin{split}
	S_{k_3}=-\frac{\rm sech^2\eta\tanh^2\eta|1-u_1^2|}{(1+2u_2\tanh\eta+(u_2^2-|1-u_1^2|)\tanh\eta)^2}\ln\frac{\rm tanh^2\eta}{(1+u_2\tanh^2\eta)^2}. 
	\end{split}
	\label{k3}
\end{equation}
Then, we arrange Eqs. \eqref{k1}, \eqref{k2}, \eqref{k3} together and define $A$ as follows, 
 \begin{equation}
 A=1+2u_2\tanh\eta+(u_2^2-|1-u_1^2|)\tanh^2\eta.
 \label{A}
 \end{equation}
Being armed with this definition, one can get $S_k$ as follows,
\begin{equation}
	\begin{split}
	&S_k=-\frac{1}{A^2}\bigg[\rm sech^2\eta[1+2u_2\tanh\eta+(u_2^2-|1-u_1^2|)\tanh^2\eta][\ln\rm sech^2-\ln(1+u_2\tanh\eta)^2]\\
	&+\rm sech^2\eta\tanh^2\eta|1-u_1^2|\ln|1-u_1^2|+|1-u_1^2|\tanh^2\eta\rm sech^2\eta[\ln\tanh^2\eta-\ln(1+u_2\tanh\eta)^2]\bigg]\\
	&=-\frac{1}{A^2}\bigg[-\ln(1+u_2\tanh\eta)^2(\rm sech^2\eta+2u_2\rm sech^2\eta\tanh\eta+u_2^2\rm sech^2\eta\tanh^2\eta )\\
	&+\rm sech^2\eta\ln\rm sech^2\eta[1+2u_2\tanh\eta+(u_2^2-|1-u_1^2|)\tanh^2\eta]+\rm sech^2\eta\tanh^2\eta|1-u_1^2|\ln|1-u_1^2|\\
	&+|1-u_1^2|\tanh^2\eta\rm sech^2\eta\ln\tanh^2\eta\bigg]\\
	&=\frac{\rm sech^2\eta}{A^2}\bigg[(1+u_2\tanh\eta)^2\ln(1+u_2\tanh\eta)^2-[(1+u_2\tanh\eta)^2-|1-u_1^2|\tanh^2\eta]\ln\rm sech^2\eta\\
	&-\tanh^2\eta|1-u_1^2|\ln|1-u_1^2|-|1-u_1^2|\tanh^2\eta\ln\tanh^2\eta\bigg].
	\end{split}
	\label{K entropy}
\end{equation}
Finally, we obtain its simplest formula.

\bibliography{mybibfile}

\end{document}